\documentclass[conference,a4paper]{IEEEtran}
\IEEEoverridecommandlockouts
\usepackage{cite}
\usepackage{amsmath,amssymb,amsfonts}
\usepackage{algorithmic}
\usepackage{algorithm}
\usepackage{graphicx}
\usepackage{textcomp}
\usepackage{xcolor}
\usepackage{booktabs}
\usepackage{url}
\def\BibTeX{{\rm B\kern-.05em{\sc i\kern-.025em b}\kern-.08em
    T\kern-.1667em\lower.7ex\hbox{E}\kern-.125emX}}

\usepackage{eso-pic}
\begin{document}
\AddToShipoutPictureBG*{%
  \AtPageUpperLeft{\put(0,-28){\makebox[\paperwidth]{\small\itshape Published in Proc. 2026 IEEE 4th International Conference on Artificial Intelligence, Blockchain, and Internet of Things (AIBThings).}}}%
  \AtPageLowerLeft{\put(0,18){\makebox[\paperwidth]{\parbox[b]{\textwidth}{\scriptsize\centering \copyright~2026 IEEE. Personal use of this material is permitted. Permission from IEEE must be obtained for all other uses, in any current or future media, including reprinting/republishing this material for advertising or promotional purposes, creating new collective works, for resale or redistribution to servers or lists, or reuse of any copyrighted component of this work in other works.\\ DOI: \url{https://doi.org/10.1109/AIBThings71029.2026.11771430}}}}}}

\title{Device-Tier Heterogeneity in Federated\\ Health-Risk Scoring: A Measured Study\\ on Consumer Edge Nodes}

\author{\IEEEauthorblockN{Eli Yune}
\IEEEauthorblockA{\textit{Balkeum Labs} \\
Seongnam, South Korea \\
eli.yune@balkeumlabs.ai}
\and
\IEEEauthorblockN{Umer Majeed}
\IEEEauthorblockA{\textit{Balkeum Labs} \\
Seongnam, South Korea \\
umer.majeed@balkeumlabs.ai}
\and
\IEEEauthorblockN{Mahesh Balan}
\IEEEauthorblockA{\textit{Claremont Graduate University} \\
Claremont, CA, USA \\
balan.mahesh@cgu.edu}
}

\maketitle

\begin{abstract}
Consumer health platforms increasingly run federated learning on home edge
nodes so that raw clinical data never leaves the household. In any real
deployment, however, households own \emph{different} sensing hardware---a basic
home owns a scale and a blood-pressure cuff, a premium home adds body-composition
and continuous-glucose sensing---so different clients can physically measure
different features. This \emph{device-tier feature heterogeneity} violates the
shared-feature-space assumption of federated averaging, yet it is largely
unstudied next to the usual label/quantity non-IID setting. Using only real,
public survey data (U.S.\ NHANES, $n\!\approx\!9{,}360$) and a CPU-only consumer
machine, we show that the common workaround---training on the features
\emph{every} home shares---discards $15.6$ AUROC points for diabetes-risk
scoring. A simple tier-aware federated averaging rule, which averages each weight
only over the homes whose devices observe that feature, recovers $13.9$ of those
points, lands within $1.6$ points of a centralized upper bound, and converges in
$2$ rounds instead of $26$---although only one quarter of homes own the
top-tier sensor. We report the on-device cost (wall-clock, package energy,
uplink bytes) measured on the node itself, and show that the per-tier marginal
contribution forms a monotone reward ladder suitable for an on-chain incentive
layer. Code and data scripts are public.
\end{abstract}


\section{Introduction}
Remote health monitoring is moving from the cloud to consumer-owned edge nodes:
a small on-premise device aggregates wearables, home medical instruments and
electronic health records, runs risk models locally, and transmits only derived
indicators, so that protected health data is bounded by the home
network~\cite{ref:rieke}. Federated learning (FL)~\cite{ref:fedavg}
is the natural training substrate---model parameters, never raw observations, are
exchanged---and secure aggregation~\cite{ref:secagg} hides individual updates.

Most FL research studies statistical heterogeneity: clients differ in
\emph{how much} data they hold and in their \emph{label} distributions
(quantity/label non-IID)~\cite{ref:noniid}. A consumer hardware deployment adds a
different and underexamined axis. Devices are sold in tiers, and the tier a
household buys determines which biometrics it can measure at all. A
\emph{Core} home with a smart scale and a blood-pressure cuff cannot produce a
glucose reading; a \emph{Total} home with a continuous glucose monitor (CGM) can.
Each client therefore observes a \emph{subset} of the global feature space, and
the subset is tied to hardware, not to chance. We call this \emph{device-tier
feature heterogeneity}.

This breaks the shared-feature-space assumption underlying federated averaging.
The usual reactions are unsatisfying. Restricting the model to the features
\emph{every} client shares (the lowest common denominator) throws away exactly
the high-value signals that the premium tiers were bought to provide. Naively
imputing the missing features to a constant and averaging over all clients lets
homes that never measured a feature dilute the few that did. Neither is
principled, and---as we show---the gap between them is large.

We make four contributions.
\begin{itemize}
\item We formalize device-tier feature heterogeneity, grounding the tiers in a
real tiered consumer-health product (Section~\ref{sec:system}).
\item We give a \emph{tier-aware} federated averaging rule: each model weight is
aggregated with per-feature participation weighting, i.e.\ only over the homes
whose devices actually observe that feature (Section~\ref{sec:method}).
\item We evaluate end-to-end on \emph{real} public data (NHANES) on a single
consumer CPU, reporting accuracy, convergence, communication and on-device
energy that we measure on the node itself (Section~\ref{sec:eval}).
\item We show the learned per-tier contribution forms a monotone reward ladder,
connecting model utility to an on-chain incentive layer
(Section~\ref{sec:contrib}).
\end{itemize}
The study uses only real, public data---no synthetic records and no perturbation
of measured values---and runs entirely on a consumer CPU with no GPU. All code
and data scripts are released.\footnote{\url{https://github.com/balkeumlabs/FedHet}}

\section{Related Work}
\textbf{Federated learning and non-IID data.} FedAvg~\cite{ref:fedavg}
coordinates training across decentralized clients. Heterogeneity is typically
framed as label/quantity skew, which degrades convergence and
accuracy~\cite{ref:noniid,ref:fedprox}; secure aggregation~\cite{ref:secagg}
protects updates. These works assume a common feature space across clients.

\textbf{Heterogeneous and partial features.} Vertical FL partitions
\emph{different features of the same records} across organizations~\cite{ref:vfl}.
Our setting is horizontal---each home holds different records---but with
heterogeneous feature \emph{availability} per client, closer to learning with
block-wise missing modalities~\cite{ref:hetero}. We address it with a minimal
change to the aggregation step rather than a new architecture.

\textbf{FL in healthcare and at the edge.} Federated training is
production-relevant in medicine~\cite{ref:rieke}, and quantization and adapter
methods increasingly enable on-premise, CPU-class deployment. We complement that
line by studying the data-side consequence of heterogeneous consumer hardware.

\textbf{Incentives and blockchain in FL.} Reliable participation needs
incentives~\cite{ref:kang}. The FLAI Protocol~\cite{ref:flai} couples
trusted-execution aggregation with sparse peer-evaluation contribution scoring
and on-chain settlement. We use its reward logic analytically: our per-tier
contribution scores instantiate the value signal such a layer would settle.

\section{System Model and Problem}
\label{sec:system}
\textbf{Tiered edge node.} We consider a consumer health edge node that ingests
home device streams into local records and trains on-device.
Hardware ships in three tiers; each added device contributes additional
measurable features (Table~\ref{tab:tiers}). Because NHANES contains no CGM
traces, the T3 glycemic feature is laboratory HbA1c, used as a proxy for the
CGM-derived estimate (GMI) that a deployed node would compute. A small \emph{profile}
(age, sex, smoking status) is entered once at setup and is always available.

\begin{table}[t]
\caption{Device tiers and the features each contributes. Higher tiers are
cumulative. SBP/DBP: systolic/diastolic blood pressure.}
\label{tab:tiers}
\centering
\footnotesize
\setlength{\tabcolsep}{4pt}
\begin{tabular}{@{}lll@{}}
\toprule
Tier & Device added & Features added \\
\midrule
\textit{profile} & (app setup) & age, sex, smoker \\
T1 Core & scale $+$ BP cuff & BMI, weight, SBP, DBP, pulse \\
T2 Advanced & $+$ body-comp.\ scale & waist circumference \\
T3 Total & $+$ CGM & HbA1c (glycemic proxy)$^{\dagger}$ \\
\bottomrule
\multicolumn{3}{@{}p{\linewidth}@{}}{\scriptsize $^{\dagger}$A CGM measures
interstitial glucose, not HbA1c (a laboratory assay). Laboratory HbA1c stands in
for CGM-derived glycemic information, e.g., the glucose management indicator
(GMI).}
\end{tabular}
\end{table}

\textbf{Federation.} The federation is a set of $N$ home nodes and a server.
Home $i$ holds a private, non-IID shard $\mathcal{D}_i$ of records and owns a
tier $t_i\!\in\!\{1,2,3\}$, which fixes an observation mask
$\mathbf{m}_i\!\in\!\{0,1\}^{D}$ over the global feature superset of size $D$:
$m_{i,j}\!=\!1$ iff a device in tier $t_i$ measures feature $j$. Let
$S_i=\{j:m_{i,j}=1\}$.

\textbf{Privacy.} Only model weights and aggregate statistics ever leave a home;
under secure multi-party computation the server observes only the masked sum, not
any individual update~\cite{ref:secagg}. This privacy substrate is assumed, not
the focus of this paper.

\textbf{Problem.} Train a single global model over the feature superset
$\{1,\dots,D\}$ when each client $i$ can compute gradients only for the features
in $S_i$. Standard FedAvg assumes $S_i\!=\!\{1,\dots,D\}$ for all $i$; here the
$S_i$ differ by hardware.

\section{Tier-Aware Federated Learning}
\label{sec:method}
\textbf{Model.} We use logistic regression over the standardized feature
superset, with weights $\mathbf{w}\!\in\!\mathbb{R}^{D}$ and bias $b$. A linear
model is deliberate: it federates cleanly, trains in milliseconds on a CPU, and
lets us mask individual features exactly to the device a home owns. Probabilities
are $\sigma(\mathbf{w}^\top\mathbf{x}+b)$.

\textbf{Federated standardization.} Per-feature mean and variance are aggregated
across the homes that observe each feature (counts and sums only, compatible with
secure aggregation); unobserved columns are set to zero, i.e.\ imputed to the
feature mean after standardization.

\textbf{Masked local update.} Client $i$ runs full-batch gradient descent on its
shard. The logistic gradient is masked so a home never moves a weight it cannot
inform:
\begin{equation}
\mathbf{g}_i = \big(\mathbf{X}_i^\top(\sigma(\mathbf{X}_i\mathbf{w}+b)-\mathbf{y}_i)/n_i + \lambda\mathbf{w}\big)\odot \mathbf{m}_i ,
\end{equation}
with $n_i=|\mathcal{D}_i|$; weights for $j\notin S_i$ stay at the broadcast value.

\textbf{Tier-aware aggregation.} The key step is to aggregate each weight only
over the homes that observe its feature, weighted by shard size:
\begin{equation}
w_j^{(t+1)} = \frac{\sum_{i:\, j\in S_i} n_i\, w_{i,j}^{(t)}}{\sum_{i:\, j\in S_i} n_i},
\qquad
b^{(t+1)} = \frac{\sum_i n_i\, b_i^{(t)}}{\sum_i n_i}.
\label{eq:agg}
\end{equation}
We contrast two baselines that share the same model and training loop:
\emph{Intersection FL} restricts the model to features common to all homes
(profile $+$ T1) and runs ordinary FedAvg; \emph{Naive union FL} keeps the full
superset but averages every weight over \emph{all} homes (the denominator in
\eqref{eq:agg} becomes $\sum_i n_i$), so non-observers dilute it. Algorithm~1
summarizes the procedure.

\begin{algorithm}[t]
\caption{Tier-aware federated averaging}
\label{alg}
\begin{algorithmic}[1]
\STATE initialize $\mathbf{w}\!\gets\!\mathbf{0}$, $b\!\gets\!0$; compute federated scaler
\FOR{round $t = 1$ to $R$}
  \FOR{each selected home $i$ (in parallel)}
    \STATE $(\mathbf{w}_i,b_i)\gets$ local GD from $(\mathbf{w},b)$ with gradient mask $\mathbf{m}_i$
  \ENDFOR
  \STATE update each $w_j$ over homes with $j\in S_i$ via \eqref{eq:agg}; update $b$
\ENDFOR
\end{algorithmic}
\end{algorithm}

\section{Experimental Evaluation}
\label{sec:eval}
\subsection{Setup}
\textbf{Data.} We use NHANES~\cite{ref:nhanes} 2015--2016 and 2017--2018 (CDC, fully public). Adults
($\geq\!20$\,y) with a valid diabetes response and complete values for every
modeled feature are kept (complete-case; $n=9{,}360$; diabetes prevalence
$15.0\%$). Keeping only complete records means device-tier heterogeneity is
imposed purely by \emph{masking} the features a home's devices cannot measure,
never by altering or fabricating any value. The label is physician-diagnosed
diabetes (NHANES \texttt{DIQ010})---a genuine condition label, not derived from
any feature.

\textbf{Federation.} We hold out $20\%$ of records as a common test set and
partition the rest into $N=60$ homes with a Dirichlet($\alpha\!=\!0.5$) label
split (non-IID); shard sizes range $7$--$752$. Tiers are assigned with mix
$50/30/20\%$ (T1/T2/T3), mirroring adoption, giving $25/19/15$ homes after empty
shards are dropped---so only one quarter of homes own the CGM.

\textbf{Model and platform.} Logistic regression, $R=40$ rounds, $5$ local
epochs, learning rate $0.5$, $\ell_2=10^{-4}$, full participation. Everything
runs on a consumer desktop (AMD Ryzen~5 5500GT, 6C/12T, 62\,GB RAM, Ubuntu
24.04); the on-board GPU is unused. Package energy is read from the Linux RAPL
interface. Runs are seeded and reproducible from the released scripts.

\subsection{Accuracy under heterogeneity}
\begin{figure}[t]
\centering
\includegraphics[width=\columnwidth]{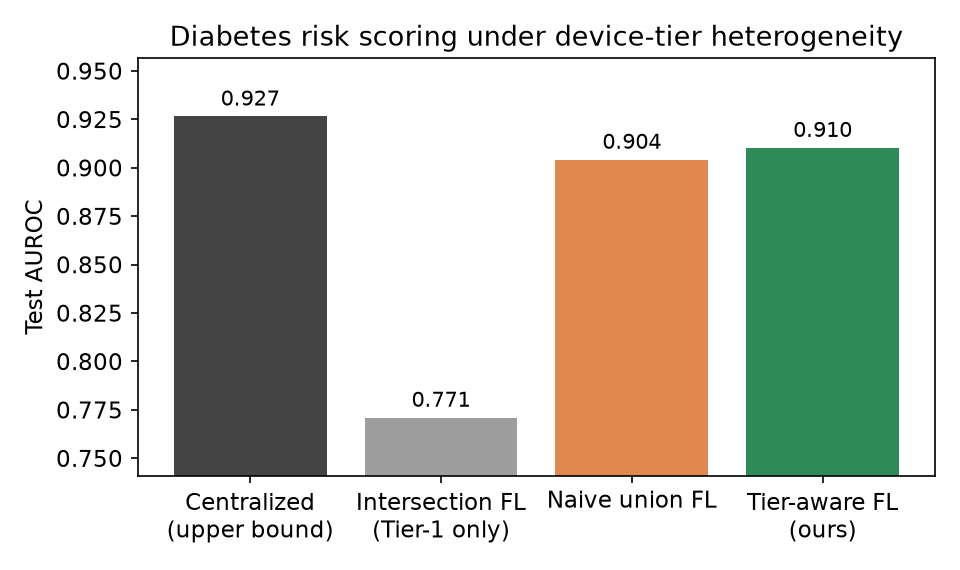}
\caption{Diabetes-risk AUROC under device-tier heterogeneity. Intersection FL
(lowest common denominator) loses $15.6$ points; tier-aware FL recovers nearly
all of it.}
\label{fig:auroc}
\end{figure}

Table~\ref{tab:main} and Fig.~\ref{fig:auroc} report test-set performance.
Handling heterogeneity by intersection collapses to AUROC $0.771$---$15.6$ points
below the centralized upper bound ($0.927$)---because it discards the glycemic
and body-composition signals the premium tiers provide. Tier-aware FL reaches
$0.910$, within $1.6$ points of centralized and $+13.9$ points over
intersection, despite only a quarter of homes owning the CGM. It also edges naive
union ($0.904$) on every metric. Notably, the profile$+$T1$+$T2 configuration
($\approx\!0.78$) matches published \emph{non-invasive} diabetes risk
scores~\cite{ref:findrisc}, while the glycemic signal of the top tier (HbA1c, our proxy for CGM-derived
glycemia) lifts performance toward the clinical gold standard---so the tier ladder is
clinically interpretable, not an artifact.

\begin{table}[t]
\caption{Test-set performance and rounds to converge (within $0.5$ pp of own
final AUROC). Gap is versus the centralized upper bound.}
\label{tab:main}
\centering
\small
\setlength{\tabcolsep}{4pt}
\begin{tabular}{@{}lcccc@{}}
\toprule
Method & AUROC & AUPRC & Gap (pp) & Rounds \\
\midrule
Centralized (upper bnd.)         & 0.927 & 0.733 & ---    & --- \\
Intersection FL                  & 0.771 & 0.330 & $-15.6$ & 3 (stuck) \\
Naive union FL                   & 0.904 & 0.675 & $-2.2$  & 26 \\
\textbf{Tier-aware FL (ours)}    & \textbf{0.910} & \textbf{0.692} & $\mathbf{-1.6}$ & \textbf{2} \\
\bottomrule
\end{tabular}
\end{table}

\subsection{Convergence and communication}
\begin{figure}[t]
\centering
\includegraphics[width=\columnwidth]{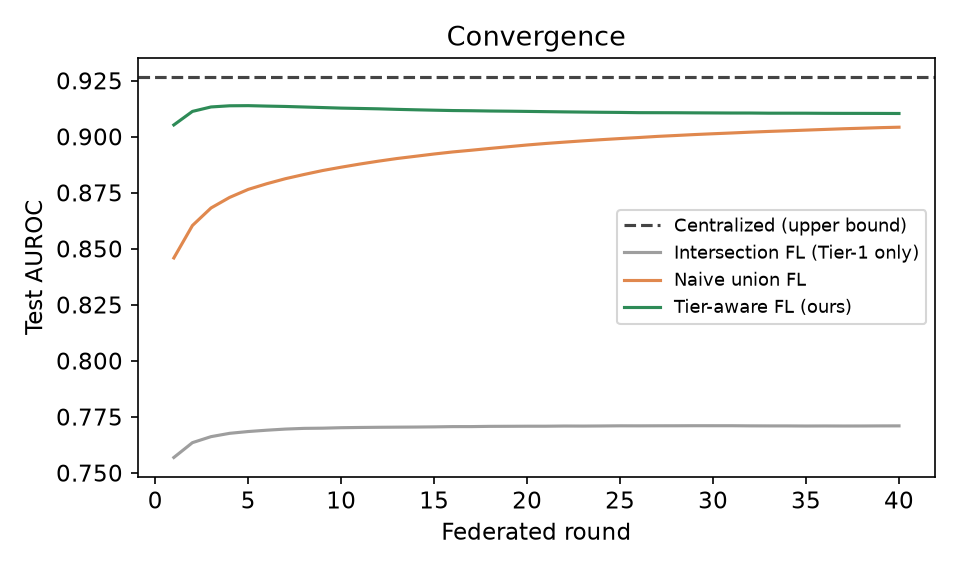}
\caption{Convergence. Tier-aware FL plateaus in $\sim\!2$ rounds; naive union
needs $\sim\!26$; intersection is capped by its missing features.}
\label{fig:conv}
\end{figure}

Fig.~\ref{fig:conv} shows the cost of heterogeneity is also temporal. Tier-aware
FL reaches its plateau in $2$ rounds (AUROC $0.905$ already at round~1), because
each weight is learned directly from its observers. Naive union needs $\approx
26$ rounds to climb to the same accuracy, as the dilution from non-observers must
be slowly overcome; intersection converges quickly but to a much lower ceiling.
At $13\times$ fewer rounds to target, tier-aware FL proportionally reduces
communication and on-device compute. The model is $11$ parameters, so each home
uploads $44$\,bytes per round and the entire run transmits $\approx\!101$\,KiB;
raw data egress is zero.

\subsection{On-device cost and scaling}
\begin{figure}[t]
\centering
\includegraphics[width=\columnwidth]{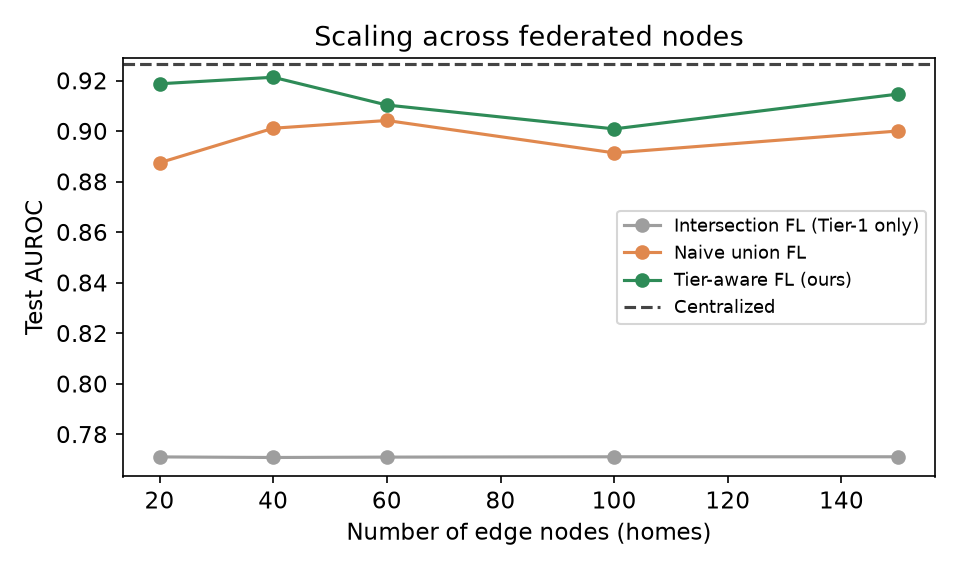}
\caption{Scaling across federated nodes. Tier-aware FL tracks the centralized
bound and dominates the baselines from $20$ to $150$ homes.}
\label{fig:scaling}
\end{figure}

Training tier-aware FL for all $40$ rounds on the node takes $0.28$\,s of
wall-clock and $6.6$\,J of measured RAPL package energy on the CPU alone---an
order of magnitude below the power envelope of a training GPU and well within a
home-appliance budget. Fig.~\ref{fig:scaling} sweeps the number of home nodes
from $20$ to $150$: tier-aware FL stays near the centralized bound throughout and
dominates both baselines, confirming the result is not an artifact of a
particular federation size.

\subsection{Tier contribution and incentive alignment}
\label{sec:contrib}
\begin{figure}[t]
\centering
\includegraphics[width=\columnwidth]{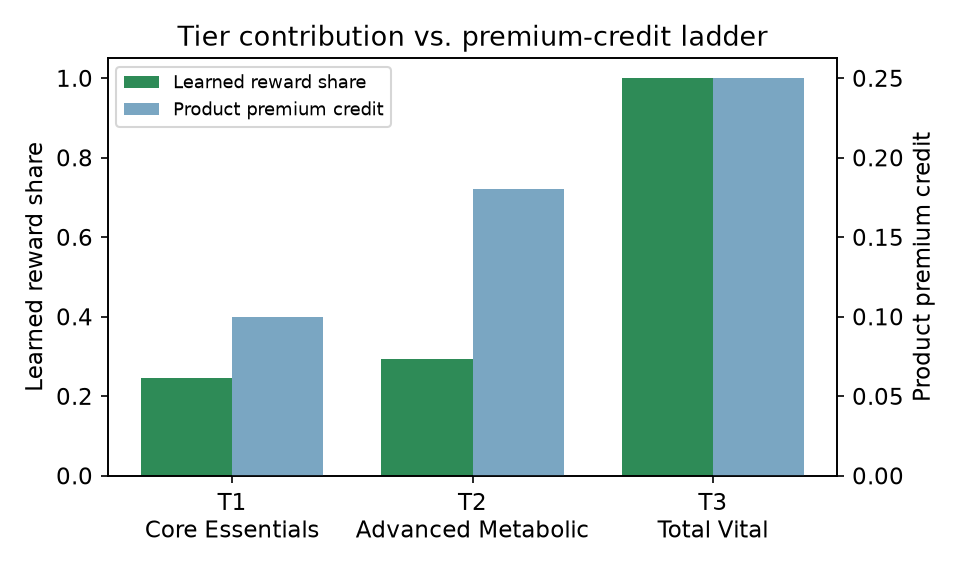}
\caption{Per-tier marginal contribution (learned reward share) against the
deployed premium-credit ladder. Both increase monotonically with tier.}
\label{fig:contrib}
\end{figure}

Because tier-aware aggregation attributes each weight to its observing homes, the
\emph{value} of a tier can be read off directly. Fig.~\ref{fig:contrib} reports
the marginal AUROC gain of each tier over the profile baseline ($0.721$):
T1 $+5.0$, T2 $+1.0$, and T3 (CGM) $+14.6$ points. The T3 gain should not be
over-read: the label is physician-diagnosed diabetes and HbA1c${}\geq{}6.5\%$ is
itself a diagnostic criterion, so part of this gain reflects the feature's
overlap with how the label is defined, not only the value of the sensor
(Section~\ref{sec:limits}). Normalizing the cumulative
gains yields a monotone reward share ($0.24/0.29/1.00$) that tracks the product's
deployed premium-credit ladder ($10/18/25\%$): homes that contribute more
predictive signal earn more. This is exactly the value signal an on-chain
incentive layer such as FLAI~\cite{ref:flai} would settle; here we compute it
analytically rather than deploying a chain.

\section{Discussion and Limitations}
\label{sec:limits}
The findings are honest about scope. NHANES is cross-sectional, so this is risk
\emph{stratification}, not incident prediction; homes are simulated by
partitioning real records rather than collected in situ. We use a linear model by
design---to isolate the aggregation effect and stay CPU-only---and deeper models
are future work. HbA1c is near-diagnostic for diabetes (HbA1c${}\geq{}6.5\%$ is a
diagnostic criterion), and a CGM does not measure it directly: we use laboratory
HbA1c as a proxy for CGM-derived glycemic information (GMI), so the T3 gain is an
optimistic estimate of what a CGM tier contributes. We note that the non-invasive tiers alone already match published risk
scores~\cite{ref:findrisc}. We deliberately excluded a medication-count feature
that, for a prevalence label, is diagnosis-adjacent and would inflate the top
tier. The incentive layer is analytical; the privacy substrate (SMPC) is assumed.
Results are reported on a single dataset and one consumer machine.

\section{Conclusion}
Consumer health hardware ships in tiers, so federated clients observe different
features---a heterogeneity axis distinct from the usual label skew. Handling it
naively is costly: training on the shared feature subset loses $15.6$ AUROC
points for diabetes-risk scoring. A one-line change to the aggregation
rule---averaging each weight only over the homes whose devices observe it---
recovers $13.9$ of those points, lands within $1.6$ of a centralized bound, and
converges $13\times$ faster, all measured on a consumer CPU with zero raw-data
egress. The same per-feature attribution yields a monotone contribution ladder
ready for an on-chain incentive layer. Tier-aware aggregation is a small,
deployable step toward federated health intelligence that respects the
heterogeneous hardware households actually own.

\sloppy

\end{document}